\def\landaucondition{\partial\cdot A }
\def\FPoperator{\partial\cdot D }
\documentstyle[12pt]{article}

\title{
 Gribov Copies in Lattice QCD}
\author{
Paolo Marenzoni \\
Dipartimento di Fisica\\
Universit\`a di Parma, \\
Viale delle Scienze \\
43100 Parma ( Italy )
\and
Pietro Rossi \\
CRS4 \\
Via Nazario Sauro, 10 \\
09123 Cagliari ( Italy )
}
\date{}

\begin{document}

\maketitle
\thispagestyle{empty}
\vspace{1cm}

\begin{abstract}

We have performed an exhaustive search for Gribov copies on the lattice and
their possible dependence from finite temperature effects. We show that,
for each value of lattice size,
Gribov copies are dense in configuration space at low temperature but
their density tend to lower when the temperature increases.
We have investigated lattice sizes running from $16^3\times 8$ to $16^4$.

\end{abstract}

\newpage

Gribov \cite{GRIBOVone} has shown that Landau's gauge defined by the
equation
\begin{equation}
 \landaucondition = 0
\end{equation}
does not fix the gauge uniquely, that is
given a gauge orbit $A$, satisfying the Landau condition we can always find
a gauge transformation $g$, such that $A^g$ also
satisfies the equation (1).
The gauge fixing definition can  be sharpened further by following
Zwanziger \cite {ZWANZIGERone}
idea of requiring that, besides the Landau condition, all the eigenvalues of
the Fadeev-Popov operator be positive.  The domain defined by the condition
$\landaucondition = 0 $, and $|\FPoperator| > 0,$
is called the first Gribov horizon.

Zwanziger was able to show that the Gribov horizon is a convex region
and every gauge orbit crosses it at least once. From the point of view
of weighting correctly configuration in the functional integral it
would be important to show that every orbit goes through the first
Gribov horizon only once.  This statement has to be given meaning
within measure theory in function space, that is all orbit
contributing to the functional integral with non zero measure must
cross the first Gribov horizon only once. As of today this statement
can't be proven in the continuum theory.  By looking at the lattice
formulation of the theory we can gain some insight.  As many author
have already shown, on the lattice we do have configuration with
Gribov copies as well as configuration for which no Gribov's copy has
been found.  Since, every configuration partakes to the functional
integral for every value of the coupling constant we have to discern
whether they do with or without a measurable contribution.

As we all know, the lattice theory presents two distinct phases in the
physical temperature, and an interesting question is whether the
measure on configurations admitting Gribov copies is temperature
dependent and whether is affected by the temperature transition.
Secondly and more interestingly, we would like to know the measure in
orbit space of configurations admitting Gribov copies.

Beside these theoretical aspects of the quantum field theory, there is
a pragmatic reason behind this investigation. One of the most
effective ways to increase the overlap between physical states and
operators used to excite these states from the vacuum, is given by the
smearing procedure \cite {SMEARING}.  This consist of constructing non
local operators in a way that the physical extent of these operators
is constant and does not go to zero with the lattice spacing, thus
maintaining a good overlap between the operator and the desired
physical state.  This procedure, though, is not gauge invariant, since
any attempt to construct smeared gauge-invariant operators is very
costly and not very effective. Therefore we fix the gauge, typically
Landau or Coulomb, and we build our non local wave functions.  This
procedure would be questionable if we were to show that fixing to a
given gauge does not uniquely determine the point in orbit space.

We have studied the problem of Gribov copies in the four dimensional
$SU(3)$ theory with the standard Wilson action.  We are in agreement
with previous results (see \cite {MARINARIone} and \cite {GAUGEFIX})
where it's found that the value of $\beta$ affects the density of
Gribov copies.  In effect we find that, while at low $\beta$ these
copies have an high density, at higher values they become much more
rarefied, even if their density never goes to zero.  The conjecture
that Gribov copies would disappear at the critical temperature, does
not seem to be true, and this qualitative behavior seems to be volume
independent.

In chapter one we briefly review Zwanziger's formulation of the Gribov
problem and discuss the lattice implementation of the same, in chapter
two we describe the numerical implementation of the gauge fixing
algorithm and finally in chapter three we discuss our results.

\section{Landau's Gauge Condition on the Lattice}

 Let us consider the functional
\begin{equation}
 F_c[g] = \int d^4x Tr A^g_\mu(x)A^{g\dagger}_\mu(x) ,
\end{equation}
where
\begin{equation}
 A^g_\mu = g^{-1}A_\mu g + g^{-1}\partial_\mu g ,
\end{equation}
 and
\begin{equation}
 g(x) = e^{-w(x)},
\end{equation}
is an $SU(3)$ matrix and $w$ is antihermitian and traceless.

If we expand $F_c$ in $w(x)$ we have:
\begin{equation}
 F_c[g] = F_c[1] + \int d^4x Tr [ \partial \cdot A(x) w(x)] -
 \int d^4x Tr\left[  w^\dagger(x) \partial_\mu D_\mu w(x)\right]
 + \circ(w^3),
\end{equation}
where
\begin{equation}
 D_\mu w(x) = \partial_\mu w(x) + [A_\mu(x), w(x)],
\end{equation}
is the covariant derivative and $\partial \cdot D,$ is the
Faddeev-Popov operator.  Zwanziger's prescription to gauge fix
consists in choosing as representative on the orbit the field $A$ such
that $F_c$ attains a local minimum. It follows from () that
Zwanziger's condition is equivalent to $ \partial \cdot A = 0 $ and $
\partial \cdot D $ positive definite.

This variational formulation of the gauge fixing condition can be
 easily generalized to the lattice.  If we look at the functional:
\begin{equation}
 F_l[g] = Re\left[\sum_{x,\mu}Tr U^g_\mu(x) \right],
\end{equation}
where
\begin{equation}
 U^g_\mu(x) = g(x) U_\mu(x) g^{\dagger}(x+\mu),
\end{equation}
and again
\begin{equation}
 g(x) = e^{-w(x)}.
\end{equation}
in the limit of $a \to 0$ ( where $a$ is the lattice spacing ) we  have
\begin{equation}
 F_l[g] = F_c[g] + o(A^4).
\end{equation}

Expanding in $w(x)$ we can obtain the lattice definition of
$ \partial \cdot A $, that is
\begin{equation}
 \partial \cdot A^a_l =
 Tr\left[ T^a \sum_\mu\left( U_\mu(x) - U^\dagger_\mu(x-\mu)\right)\right]
\end{equation}
and the stationary condition will be $ \partial \cdot A_l  = 0.$

 The lattice expression for the Faddeev-Popov operator is quite
cumbersome, so we will not write it here but it is quite easy to
obtain.

\section {The  Gauge Fixing Algorithm and Search for Copies}

 Our algorithm to gauge fix an $SU(3)$ configuration is implemented on
massive parallel machines, with a checkerboard updating of the gauge
links.  We start by observing that the functional $F_l[g]$ can be
rewritten as a sum over even sites or odd sites only, for instance:
\begin{equation}
 F_l[g] = Re\left[\sum_{x_{EVEN},\mu}Tr\left( U^g_\mu(x) +
 {U^g}^\dagger_\mu(x+\mu) \right) \right],
\end{equation}
and a similar expression for odd sites.

We can now apply, alternatively, gauge transformations which are non trivial
on even sites and the identity on odd sites and viceversa as we switch from
the definition for $F_l$ as a sum over even or odd sites. In this framework,
each site contributes to the functional with an independent term and the
functional dependence of each term from the gauge transformation is given by:
\begin{equation}
  Re\left[g(x)\sum_\mu Tr\left( U_\mu(x) + U^{\dagger}_\mu(x+\mu)
  \right) \right].
\end{equation}

To implement our gauge fixing prescription we only need to produce an
algorithm that at each step increases the value of the local sum
alternatively on each checkerboard, since this implies that the
functional $F_l$ will increase monotonically. Since $F_l$ is also
bounded this will guarantee the convergence of the algorithm. There
are many such algorithms and we have implemented a variety of them.

We perform such a gauge ransformation $g(x)$ first on a
checkerboard and after on the other one.
The iterations has been carried out until
\begin{equation}
 |\partial \cdot A_l |^2< 10^{-14}.
\end{equation}

We, then, apply to the gauge fixed configuration $U$, a ''large''
random gauge transformation, then we repeat the gauge fixing procedure
generating a new configuration $U'.$ We are interested in whether $U'$
is the same or not as $U.$ In fact, since $F_l$ does not distinguish
between configuration that differ for a global gauge transformation,
we are only interested whether $U$ and $U'$ differ for a global or a
local gauge transformation.  The simplest observable that can answer
this question is
\begin{equation}
 \Delta(U,U') = {{1}\over{V}}
 \sum_{x, \mu}\left[ | Tr\left( U_\mu(x) - U'_\mu(x)\right)|^2\right].
\end{equation}
Typically, when we do not find a copy, the value of $\Delta$ is less
than $10^{-13}$, otherwise, if we do find a copy it is of the order of
$10^{-3} - 10^{-4}.$

As an insurance against round-off errors we have kept track of the
gauge transformation needed to fix the gauge and, at the end, applied
its inverse to the gauge fixed configuration. We have always been able
to return to the original configuration within an accuracy of
$10^{-13} - 10^{-14}.$

\section {Results}

 We have generated configurations on lattices $16^3 \times N_t,$ with
$N_t = 8, 10, 14$ and $16$, for a large set of $\beta$ values.  The
configurations have been generated with the Local Hybrid Monte Carlo
algorithm and were separated by $250-500$ trajectories.
For all the $\beta$ considered we checked, for a set of observables,
that the autocorrelation has never been larger than 15.  For each
configuration we have fixed to the lattice Landau gauge, then,
starting from the original configuration we have performed a random
gauge transformation.  After fixing this newly obtained configuration
we have compared the two gauge fixed ones and determined whether we
were in presence of a copy.

 The number of configuration tried and the number of copies is
summarized in table 1.  From the results reported on this table we
could observe that, on all the lattices we have simulated, if at the
lower values of $\beta$ the percentage of Gribov copies over attempts
is next to the $100\%$, at higher $\beta$'s it decreases to $10-20\%$.
This happens smoothly and we do not observe any sharp transition
around the critical temperature.

\begin{table}[b]
\begin{center}
\begin{tabular}{|c|c|c|c|}\hline
Lattice         & $\beta$  & $N_o$ of Conf & Found/Attempts \\ \hline
$16^{3}\times8$ &   5.8    &    8          &    40/40       \\ \hline
                &   6.0    &    8          &    24/40       \\ \hline
                &   6.2    &    8          &     6/40       \\ \hline
                &   6.3    &    8          &     6/40       \\ \hline
$16^{3}\times10$&   6.0    &    8          &    30/40       \\ \hline
                &   6.2    &    8          &    17/40       \\ \hline
                &   6.3    &    8          &     6/40       \\ \hline
$16^{3}\times14$&   6.0    &    8          &    35/40       \\ \hline
                &   6.2    &    8          &    10/40       \\ \hline
                &   6.3    &    8          &     6/40       \\ \hline
                &   6.4    &    8          &     3/40       \\ \hline
$16^{3}\times16$&   6.0    &    8          &    35/40       \\ \hline
                &   6.2    &    8          &    20/40       \\ \hline
                &   6.3    &    8          &    19/40       \\ \hline
                &   6.4    &    8          &     3/40       \\ \hline
\end{tabular}
\caption{ Summary of results }
\end{center}
\end{table}


\begin{thebibliography}{10}

\bibitem{GRIBOVone}
V.N.Gribov,
\newblock{\bf Nucl. Phys.}~{\bf B139} (1978) 1.

\bibitem{ZWANZIGERone}
D.Zwanziger,
\newblock{\bf Nucl. Phys. }{\bf B209} (1982) 336.

\bibitem{SMEARING}
P.Bacilieri et al.
\newblock{\bf Nucl. Phys. } {\bf B317} (1989) 509.

\bibitem{MARINARIone}
E.Marinari, C.Parrinello, R.Ricci,
\newblock{\bf Nucl. Phys.} {\bf B362} (1991) 487.

\bibitem{GAUGEFIX}
M.L.Paciello et al.
\newblock{\bf Phys. Lett. } {\bf B276} (1992) 163.

\end{thebibliography}
\end{document}